\def\section{\setcounter{equation}{0}
\@startsection {section}{1}{\z@}{-3.5ex plus -1ex minus
 -.2ex}{2.3ex plus .2ex}{\large\bf}}

\def\ageq{\vcenter{\vbox{\hbox{$\buildrel > \over \sim$}}}}
\def\square#1{\mathop{\mkern0.5\thinmuskip\vbox{\hrule\hbox{\vrule\hskip#1
              \vrule height#1 width 0pt \vrule} \hrule} \mkern0.5\thinmuskip}}
\documentstyle [12pt] {article}

\newcommand{\ns}{\vfill\eject}
\newcommand{\be}{\begin{equation}}
\newcommand{\ee}{\end{equation}}
\newcommand{\bes}{\begin{eqnarray}}
\newcommand{\ees}{\end{eqnarray}}

\newcommand{\nummer}[1]{\hfill #1 \par}
\newcommand{\monat}[1]{\vspace{-14pt}\hfill #1
                       \par \vspace*{1 cm}}
\newcommand{\titel}[1]{{\renewcommand{\thefootnote}{\fnsymbol{footnote}}
                       \Large\bf\vskip 0 true cm
                       \begin{center}#1\end{center}
                       \setcounter{footnote}{0}}
                       \normalsize\vskip 1.2 true cm}
\newcommand{\autor}[1]{{
                       \begin{center} {\large #1 }\end{center}}
                       \setcounter{footnote}{0}}
\newcommand{\adresse}[1]{\vspace*{-1.1 true cm}\begin{center} {\it #1 } 
                         \end{center}
                         \vskip 0.5cm}

\begin{document}
\thispagestyle{empty}
\nummer{WU-B 93-07}
\monat{September 1993}
\titel{OCTET-BARYON FORM FACTORS \\ IN THE DIQUARK MODEL\footnote{Sponsored in
part by the Bundesministerium f\"ur Forschung and  Technologie, FRG \\under contract
number 06 Wu 765}} \autor{R. Jakob, 
P. Kroll\footnote{E-mail: kroll@wpts0.physik.uni-wuppertal.de}, and M. Sch\"urmann}
\adresse{Fachbereich Physik, Universit\"at Wuppertal,\\  D-42097 Wuppertal, Germany}
\autor{W. Schweiger}
\adresse{Institut f\"ur Theoretische Physik, Universit\"at Graz, \\ A-8010 Graz, Austria}
{\bf Abstract:\\} {\it We present an alternative parameterization of the quark-diquark
model of baryons which particularly takes care of the most recent proton electric
form-factor data from the E136 experiment at SLAC. In addition to electromagnetic form
factors of the nucleon, for which good agreement with data is achieved, we discuss the
weak axial vector form factor of the nucleon as well as electromagnetic form factors of
$\Lambda$ and $\Sigma$ hyperons. Technical advance in calculating the pertinent analytic
expressions within perturbative quantum chromodynamics is gained by formulating the
wave function of the quark-diquark system in a covariant way. Finally, we also comment
on the influence of Sudakov corrections within the scope of the diquark model.}

\ns
\renewcommand{\thefootnote}{\arabic{footnote}} 
\setcounter{footnote}{0}
\section{ INTRODUCTION }
In order to deal with exclusive hadronic reactions at moderately large momentum
transfers ($Q^2 \ageq 4 \hbox{GeV}^2$) we have advocated a model in which baryons are
treated as quark-diquark systems \cite{kro:91a}-\cite{kro:93}. This model
incorporates scalar (S) and vector (V) diquarks and relies on factorization of
short- and long-distance dynamics -- i.e., a hadronic amplitude is expressed as a
convolution of a hard-scattering amplitude $\widehat{T}$, calculable within
perturbative quantum chromodynamics (QCD), with distribution amplitudes (DAs)
$\Phi$ which contain the (non-perturbative) binding of the hadronic constituents.
Hence, the diquark model can be considered as a modification of the pure quark
hard-scattering picture (HSP) \cite{lep:80} which is generally believed to be the
correct description of hard exclusive processes at asymptotically large momentum
transfers ($Q^2 \rightarrow \infty$). In fact, diquarks are only an effective way to
cope with non-perturbative effects still present in the kinematic range we are
interested in. One example for this kind of effects is the large asymmetry of the
quark momentum distribution inside the nucleon found by means of QCD sum
rules \cite{che:84}. Just such an asymmetric DA occurs to be necessary to explain the
nucleon magnetic form-factor data within the pure quark HSP. It is a clear indication
for the existence of strong two-quark correlations inside the nucleon \cite{ste:89}. 
 
The main ingredients of the diquark model are baryon DAs in terms of quarks and
diquarks, the coupling of gluons and photons to diquarks, and, in order to account for
the composite nature of diquarks, phenomenological vertex functions (diquark form
factors). The proper choice of the diquark form factors guarantees the compatibility
of the diquark model with the pure quark HSP in the limit $Q^2 \rightarrow \infty$.
With a common set of parameters specifying the diquarks and process independent DAs 
for the involved hadrons a good description of electromagnetic nucleon form factors
(in the space-like region) \cite{kro:91a}, electroexcitation of $\Delta$ and  $S_{11}$
resonances \cite{kro:92}, Compton scattering off protons ($\gamma p \rightarrow \gamma
p$) \cite{kro:91b}, and photoproduction of Kaons ($\gamma p \rightarrow K^+ \Lambda$)
\cite{kro:93}, \cite{sch:92} has been accomplished. Thereby, V diquarks prove to be
essential in describing spin effects caused by flips of the baryon helicity. These spin
effects are also of non-perturbative origin and play an important role in most
exclusive hadronic processes at experimentally accessible values of the momentum
transfer $Q^2$. They cannot be explained within the pure quark HSP, in which they are
suppressed by powers of $(\tilde{m}/Q)$ ($\tilde{m}$ denotes a mass of the order of
the hadronic mass).

The parameters of the diquark model were essentially determined by a fit to the
world data on elastic electron-nucleon scattering for $Q^2 \ageq 3.5 \hbox{GeV}^2$ (cf.
\cite{kro:91a}). With the exception of elastic electron-proton differential
cross sections \cite{arn:86}, which fix the proton magnetic form factor
$G_M^p(Q^2)$ very accurately up to $Q^2 \approx 31 \hbox{GeV}^2$, the quality and
quantity of these data was rather poor. As a consequence there still remained
ambiguities in the model parameters, in particular those for the V diquarks.
Very recently, however, better data on the electric form factor of the proton
$G_E^p(Q^2)$ in the momentum-transfer range $1.75 \hbox{GeV}^2 \leq Q^2 \leq 8.83
\hbox{GeV}^2$ have been published \cite{bos:92}. In reproducing the new data our model
falls somewhat short. It is thus one of the goals of the present paper to
reconcile the diquark-model results with experiment. As can be checked most easily
in the  Breit frame this affects primarily the parameters of the V diquarks. In that
frame the Pauli form factor $F_2^p(Q^2)$, which determines the difference between
$G_E^p(Q^2)$ and $G_M^p(Q^2)$, only contributes to helicity-flip transitions of the
proton. Such transitions, as we already mentioned, are generated by V diquarks.

An additional reason that makes a readjustment of the diquark parameters necessary is 
the introduction of covariant quark-diquark wave functions. According to the HSP,
neglecting intrinsic transverse momenta, we treat quarks and diquarks as free on-shell
particles with momenta collinear to that of the parent baryon. Therefore the
valence (quark-diquark) Fock state of a baryon can be written in a covariant way in
terms of the baryonic momentum and helicity (and of course a colour and flavour part).
The derivation of such a representation parallels the one for covariant wave functions
in the heavy quark effective theory \cite{geo:91}. The use of covariant wave functions
has many technical advantages. One works with hadronic quantities (spinors, helicities,
momenta, $\cdots$) from the beginning. The calculation of a large set of scattering
amplitudes at the constituent level, which finally have to be combined to a hadronic
amplitude, is avoided; one immediately projects onto hadronic states. Last not least,
covariant wave functions turn out to be extremely helpful in algebraic computer
programs, the use of which is unavoidable for calculating processes more complicated
than electron-nucleon scattering. Their introduction, however, slightly changes the
original analytic results for the V-diquark contributions. The modifications are
related to helicity flips of the quarks neglected in our previous work.

A further objective of this paper is to provide predictions for two types of form
factors not considered till now within the diquark model, namely the axial vector
form factor of the nucleon and electromagnetic form factors of hyperons. For the first
one there exist already a few data up to $Q^2 = 3 \hbox{GeV}^2$ \cite{kit:83}. These can be 
compared with our model results if one assumes that the usual dipole
parameterization gives the right trend for higher values of $Q^2$. On the other
hand, for hyperon form factors no data are available at present. Nevertheless, their
consideration might become of interest if plans will be realized to measure these form
factors by scattering a high-quality hyperon beam on an atomic electron target
\cite{gar:90}. We obtain results for the hyperon form factors by employing a
flavour dependent distribution amplitude for the baryon octet.

Eventually, we also want to comment on the influence of Sudakov suppression in the
HSP involving diquarks. Recently Li and Sterman \cite{li:92a}, \cite{li:92b} have
disputed assertions \cite{isg:89}, \cite{rad:91} that perturbative QCD is not
applicable at experimentally accessible momentum transfers. The authors of 
\cite{isg:89} and \cite{rad:91} have pointed out that an asymmetric DA like the one of
Ref. \cite{che:84} strongly enhances contributions from kinematic regions where
perturbative QCD does not hold. The main issue of Li\rq s and Sterman\rq s work is
that such contributions are suppressed and hence the validity of perturbative QCD is
reestablished if one takes into consideration the transverse momenta of the quarks and
radiative (Sudakov) corrections. In the light of these findings it is also
interesting to know whether a similar mechanism works for diquarks.

The paper is organized as follows. In Sect. 2 we introduce covariant quark-diquark
wave functions and briefly summarize the (Feynman) rules of the diquark model.
Sect. 3 contains the results for the electromagnetic form factors of the nucleon 
with a discussion of Sudakov corrections  and the predictions for hyperon form
factors. Sect. 4 is devoted to the axial vector form factor of the nucleon. The summary
is given in Sect. 5. 
\section{THE HARD-SCATTERING PICTURE \hspace{20.0truecm} . WITH DIQUARKS}
As we already mentioned in the introduction there are two elements entering a form
factor calculation in the framework of the HSP, namely a hard scattering amplitude
$\widehat{T}$ to be calculated in collinear approximation within perturbative QCD and
DAs $\Phi$. To leading order in $(1/Q)$ the form factors are determined by the
valence   Fock state of the considered hadron -- in our model a quark-diquark state in
case of an ordinary baryon. For the lowest lying baryon octet, assuming zero relative
orbital angular momentum between quark and diquark, the Fock state written in
a covariant fashion (omitting the colour part) reads
\be
\vert B; p, \lambda \rangle = f_S \Phi_S^B (x_1) \, \chi^B_S u( p,\lambda ) + 
   f_V \Phi_V^B (x_1) \, \chi^B_V \, (\gamma^\alpha + p^\alpha / m_B ) \, \gamma_5
   u( p,\lambda ) / \sqrt{3}\, .
\label{fock} 
\ee
$u$ is the spinor of the baryon, $p$ and $m_B$ its momentum and mass,
respectively. The two terms in (\ref{fock}) represent configurations consisting of a 
quark and either an S or a V diquark. The Lorentz-index $\alpha$ labels the components
of the V-diquark polarization vector. In the region where only a few $\hbox{GeV}$ of momentum
are transferred the mass of the baryon is not completely negligible. If it is taken into
account and the sum of the constituents\rq \  four-momenta is demanded to give the
baryon four-momentum one ends up with running masses $m_q = x_1 m_B$ and $m_D = x_2
m_B$ for the quark and diquark, respectively \footnote{$x_1$ denotes the fraction of
the baryon momentum carried by the quark ($x_2 = 1 - x_1$ that of the diquark).}. Since
this implies that quark, diquark, and baryon have the same (four) velocity, the chain
of arguments leading to covariant wave functions in the heavy quark effective theory
\cite{geo:91} can immediately be transferred to our collinear situation to prove
(\ref{fock}).

We assume an SU(6)-like spin-flavour dependence for the octet-baryon wave functions.
Hence the flavour functions $\chi$ for proton, neutron and hyperons take on the form 
\be
\chi_S^p = u S_{[u,d]} \, , \;  
\chi_V^p = \phantom{-} [u V_{\{u,d\}} -\sqrt{2} d V_{\{u,u\}}] / \sqrt{3} \, ,
\label{flav1} \ee
\be
\chi_S^n = d S_{[u,d]} \, , \;
\chi_V^n = - [d V_{\{u,d\}} -\sqrt{2} u V_{\{d,d\}}] / \sqrt{3} \, ,
\label{flav2} \ee
\be
\chi_S^{\Sigma^+} = - u S_{[u,s]} \, , \;  
\chi_V^{\Sigma^+} = \phantom{-}[u V_{\{u,s\}}-\sqrt{2} s V_{\{u,u\}}] / \sqrt{3} \, ,
\label{flav3} \ee
\be 
\chi_S^{\Sigma^-} = \phantom{-} d S_{[d,s]} \, , \;  
\chi_V^{\Sigma^-} = - [d V_{\{d,s\}}-\sqrt{2} s V_{\{d,d\}}] / \sqrt{3} \, ,  
\label{flav4} \ee 
\be
\chi_S^{\Sigma^0} = [ d S_{[u,s]} + u S_{[d,s]} ] / \sqrt{2} \, , \;  
\chi_V^{\Sigma^0} = [ 2 s V_{\{u,d\}}- d V_{\{u,s\}} - u V_{\{d,s\}}] / \sqrt{6} \, ,
\label{flav5} \ee
\be 
\chi_S^{\Lambda^0} = [ u S_{[d,s]} - d S_{[u,s]} - 2 s S_{[u,d]}] / \sqrt{6}  \, ,
\;   
\chi_V^{\Lambda^0} =  [u V_{\{d,s\}}- d V_{\{u,s\}}] / \sqrt{2} \, .  
\label{flav6} 
\ee 

The DA $\Phi_{S(V)}^B(x_1)$ is nothing else but a light-cone wave function integrated
over transverse momentum. The $r=0$ value of the corresponding configuration space
wave function is given by the constant $f_{S(V)}$. For our purposes the
phenomenological ansatz 
\be 
\phi_S^B (x_1) = N_S x_1 x_2^3 
\exp \left[ - b^2 \left( {m_q^2 / x_1} + {m_S^2 / x_2} \right) \right] \label{phis}
\ee
for the DA of the S diquark and the slightly more complicated expression
\be 
\phi_V^B (x_1) = N_V x_1 x_2^3 (1 + c_1 x + c_2 x^2)
\exp \left[ - b^2 \left( {m_q^2 / x_1} + {m_V^2 / x_2} \right) \right] \label{phiv}
\ee
for the DA of the V diquark occur to be quite appropriate. The analytic form of
these DAs originates from a nonrelativistic harmonic-oscillator wave
function \cite{hua:89}. Therefore the masses appearing in the exponentials have to be
considered as constituent masses. We take $330 \hbox{MeV}$ for light quarks, $580 \hbox{MeV}$ for
(light) diquarks, and add $150 \hbox{MeV}$ for each strange quark. The oscillator parameter
$b^2$ is taken to be $0.248 \hbox{GeV}^{-2}$, so that the full wave function gives rise to a
reasonable value of $600 \hbox{MeV}$ for the mean intrinsic transverse momentum of quarks
inside a nucleon. The \lq\lq normalization\rq\rq \ constants $N_S$ and $N_V$
are determined by the condition $\int dx_1 \, \Phi_{S (V)} (x_1) = 1$.

The Feynman diagrams contributing to the hard-scattering amplitude of baryon form
factors are displayed in Fig. 1. The gluon- and photon-diquark vertices are defined by
\bes
S \left( \begin{array}{c} g \\ \gamma \end{array} \right) S &:& 
i \left( \begin{array}{c} + g_s t^a \\ -e_0 e_S \end{array} \right) (p_1 + p_2)_\mu
\, , \label{svert} \\ & & \nonumber \\
V \left( \begin{array}{c} g \\ \gamma \end{array} \right) V &:&
i \left( \begin{array}{c} - g_s t^a \\ + e_0 e_S \end{array} \right)
[ g_{\alpha \beta} (p_1 + p_2)_\mu - \\
& &   g_{\mu \alpha} [(1+\kappa_V) p_1 - \kappa_V p_2 ]_\beta
     - g_{\mu \beta}  [(1+\kappa_V) p_2 - \kappa_V p_1 ]_\alpha ] \nonumber
\, , \label{vvert}  
\ees 
with $g_s = \sqrt{4 \pi \alpha_s}$ designating the QCD coupling constant, $\kappa_V$
the anomalous (chromo)mag\-netic 
moment of the vector diquark and $t^a (=\lambda^a/2)$ the
Gell-Mann colour matrix \footnote{The different signs in electromagnetic ($e_0 > 0$)
and strong vertices are due to the fact that the diquark is in a colour state 
belonging to the antitriplet.}. Gauge invariance also calls for contact terms
\bes
\gamma S g S & : & - 2 i e_0 e_S g_s t^a g_{\mu \nu} 
\, , \label{svert4} \\ & & \nonumber \\
\gamma V g V & : & + i e_0 e_V g_s t^a (2 g_{\mu \nu} g_{\alpha \beta}
                   - g_{\mu \beta} g_{\alpha \nu} -g_{\mu \alpha} g_{\beta \nu})
\, . \label{vvert4}
\ees

In applications of the diquark model Feynman diagrams are calculated with these rules
for point-like particles. In order to take into account the composite nature
of diquarks phenomenological vertex functions (diquark form factors) have to be
introduced. Our choice    
\bes
F_S^{(3)} (Q^2) &=& \delta_S {Q_S^2 \over Q_S^2 + Q^2} \; , 
\\
F_V^{(3)} (Q^2) &=& \delta_V \left( {Q_V^2 \over Q_V^2 + Q^2}\right)^2 \; ,
\label{form3}
\ees 
for 3-point functions and 
\be
F_S^{(n)} = a_s F_S^{(3)} (Q^2) \, , \;\;\;\; 
F_V^{(n)} = a_V F_V^{(3)} (Q^2) \left( {Q_V^2 \over Q_V^2 + Q^2}\right)^{(n-3)}
\label{form4}
\ee
for n-point functions ($n \geq 4$) ensures that in the limit $Q^2 \rightarrow
\infty$ the diquark model evolves into the pure quark HSP. The factor
$\delta_{S (V)}  =  \alpha_s (Q^2) / \alpha_s (Q^2_{S (V)})$ ($\delta_{S (V)} = 1$ for
$Q^2 \leq  Q^2_{S (V)}$) provides the correct powers of $\alpha_s (Q^2)$ for
asymptotically large $Q^2$. $\alpha_S = 12 \pi / 25 \ln (Q^2 / \Lambda_{QCD}^2 )$ is
used with $\Lambda_{QCD} = 200 {\rm \hbox{MeV}}$ and restricted to be smaller than $0.5$.
$a_S$ and $a_V$ are strength parameters which allow for the possibility of diquark
excitation and break-up in intermediate states where diquarks can be far off-shell.  
\section{ELECTROMAGNETIC FORM FACTORS}
For baryons belonging to the lowest lying baryon octet ($J^P = 1/2^+$) the matrix
elements of the electromagnetic current operator can be cast into the form
\be
\langle B; p_f, \lambda_f \vert \widehat{J}^\mu \vert B; p_i, \lambda_i \rangle =
-i e_0 \bar{u} (p_f, \lambda_f) \left[ \gamma^\mu G_M^B (Q^2) - {\kappa_B \over 2 m_B}
(p_f + p_i)^\mu F_2^B (Q^2) \right] u (p_i, \lambda_i) \, ,
\label{emcurr}
\ee
with $\kappa_B$ denoting the anomalous magnetic moment of the baryon $B$. The
magnetic form factor $G_M^B$ and the Pauli form factor $F_2^B$ are phenomenological
functions of $Q^2 = (p_f - p_i)^2$ which parameterize the electromagnetic structure of
the baryon $B$. The two other commonly used form factors are the electric form factor
$G_E^B$ and the Dirac form factor $F_1^B$. They are related to $G_M^B$ and $F_2^B$ via
$F_1^B = G_M^B - \kappa_B F_2^B$ and $G_E^B = G_M^B - \kappa_B (1 + Q^2 / (4 m^2) )
F_2^B$, respectively. 

At the constituent level the electromagnetic current is obtained as a convolution of
hard scattering amplitudes $\widehat{T}$, given by the Feynman diagrams of Fig.1, with
the pertinent DAs (\ref{phis}) and (\ref{phiv}) for S and V diquarks, respectively. The
covariant formulation (\ref{fock}) of the baryon wave functions enters the calculation
of the Feynman diagrams as far as the spinor $u (p_q, \lambda_q)$ of an external
incoming quark has to be replaced by the spinor of the baryon $u (p, \lambda)$ to which
it belongs and the polarization vector $\epsilon^\alpha (p_V, \lambda_V)$ of an external
incoming V diquark has to be replaced by the factor $(\gamma^\alpha + p^\alpha / m_B)
\gamma_5$ put in front of the baryon spinor $u (p, \lambda)$. For outgoing quarks we
have instead $\bar{u} (p, \lambda)$ and for outgoing V diquarks $\gamma_5
(\gamma^\alpha + p^\alpha / m_B)$ at the right hand side of this spinor. Thereby
it is already taken into account that the $\lambda = \pm 1/2$ helicity state of a baryon
consists of a combination of two different quark-V-diquark helicity states
\footnote{As a consequence of the collinear approximation $p_q = x_1 p$ and $p_V = x_2
p$ this combination $\propto ( u (p_q,  \lambda) \epsilon^\alpha (p_V, 0) - \sqrt{2} u
(p_q, -\lambda) \epsilon^\alpha (p_V, 2\lambda))$ can be rewritten as $ \sqrt{x_1}
(\gamma^\alpha + p^\alpha /m_B) \gamma_5 u(p, \lambda)$.}. 
Thus, only baryonic quantities show up in the analytic expressions for the Feynman
diagrams. After some commutations and subsequent use of the Dirac equation ($p \!\!
/ u(p, \lambda) = m_B u(p, \lambda)$ ) one immediately ends up with expressions for
the various contributions to the electromagnetic current which have the same covariant
structure as (\ref{emcurr}). Folding these expressions with the DAs yields
\bes
S_3 (Q^2) & = & C_F {4 \pi \over Q^2} f_S^2 \int_0^1 dx_1 dy_1 \Phi_S(y_1) 2 {\alpha_s
(\tilde{Q}^2_{22}) \over x_2 y_2 } F_S^{(3)} (\tilde{Q}^2_{22}) \Phi_S(x_1) \, , 
\label{S3} \\
S_4 (Q^2) & = & C_F {4 \pi \over Q^2} f_S^2 \int_0^1 dx_1 dy_1 \Phi_S(y_1) 2 {\alpha_s
(\tilde{Q}^2_{11}) \over x_1 y_1 } F_S^{(4)} (\tilde{Q}^2_{11} + \tilde{Q}^2_{22})
\Phi_S(x_1) \, , 
\label{S4} \\
V_3 (Q^2) & = &  C_F {4 \pi \over 9 m_B^2} f_V^2 \int_0^1 dx_1 dy_1 \Phi_V(y_1) 
{\alpha_s (\tilde{Q}^2_{22}) \over x_2 y_2 } F_V^{(3)} (\tilde{Q}^2_{22}) \Phi_V(x_1)
\kappa_V \, , 
\label{V3} \\
& & \nonumber \\
V_4 (Q^2) & = &  C_F {4 \pi \over 9 m_V^2} f_V^2 \int_0^1 dx_1 dy_1 \Phi_V(y_1) 
{\alpha_s (\tilde{Q}^2_{11}) \over x_1 y_1 } F_V^{(4)} (\tilde{Q}^2_{11} + 
\tilde{Q}^2_{22}) \Phi_V(x_1) \nonumber \\
& & \!\!\!\!\!\!\!\!\!\!\!\!
\left( \kappa_V^2 (1 + 2 x_1 + 2 y_1 + x_1 y_1) + {1 \over 2} \kappa_V (1 -
\kappa_V) (x_2 + y_2) -{3 \over 2} x_2 y_2 (1 - \kappa_V^2) \right)  \, , 
\nonumber \\ & & \label{V4} 
\ees
for the Lorentz-invariant functions in front of $\gamma^\mu$ and
\bes
V_3^f (Q^2) & = & - C_F {4 \pi \over 9 Q^2 m_B} f_V^2 \int_0^1 dx_1 dy_1 \Phi_V(y_1) 
{\alpha_s (\tilde{Q}^2_{22}) \over x_2 y_2 } F_V^{(3)} (\tilde{Q}^2_{22}) \Phi_V(x_1)
\nonumber \\ & & \phantom{- C_F {4 \pi \over 9 Q^2 m_B} f_V^2 \int_0^1}
\left( 4 \kappa_V - (x_2 + y_2) (2 + 3 \kappa_V) \right) 
\, , \nonumber \\ & & 
\label{V3f} \\
V_4^f (Q^2) & = & -  C_F {4 \pi \over 9 m_V^2 m_B} f_V^2 \int_0^1 dx_1 dy_1 \Phi_V(y_1) 
{\alpha_s (\tilde{Q}^2_{11}) \over x_1 y_1 } F_V^{(4)} (\tilde{Q}^2_{11} + 
\tilde{Q}^2_{22}) \Phi_V(x_1) \nonumber \\
& & \phantom{-  C_F {4 \pi \over 9 m_V^2 m_B} f_V^2 \int_0^1}
\left( (\kappa_V + {1 \over 2} ) (x_1 + y_1) + \kappa_V -1 \right)  \, , 
\nonumber \\ & & \label{V4f} 
\ees
for the Lorentz-invariant functions in front of $(p_f + p_i)^\mu$. $C_F = 4/3$ denotes
the colour factor and $\tilde{Q}^2_{i j} = x_i x_j Q^2$. $m_V = 580$MeV is the mass of
the V diquark. It appears in the V-diquark propagator of the 4-point diagrams.
Finally, these contributions have to be multiplied with appropriate charge and flavour
factors and summed up in order to obtain $G_M^B$ and $F_2^B$. 
\subsection{Nucleon form factors}
For nucleon form factors the pertinent combinations of scalar and vector n-point
contributions read:
\bes
G_M^p & = & e_u S_3 + e_{u d} S_4 + (e_u + 2 e_d) V_3  + (e_{u d} + 2 e_{u u}) V_4
\, , \label{gmp} \\
G_M^n & = & e_d S_3 + e_{u d} S_4 + (e_d + 2 e_u) V_3  + (e_{u d} + 2 e_{d d}) V_4
\, , \label{gmn} \\
F_2^p & = & -{2 m_p \over \kappa_p} \left( (e_u + 2 e_d) V^f_3  + (e_{u d} + 2 e_{u u})
V^f_4 \right) \, , \label{f2p} \\
F_2^n & = & -{2 m_n \over \kappa_n} \left( (e_d + 2 e_u) V^f_3  + (e_{u d} + 2 e_{d d})
V^f_4 \right) \label{f2n} \, .
\ees
$e_q$ and $e_{q_1 q_2}$ are the electric charges of the quarks and diquarks
(consisting of quarks $q_1$ and $q_2$) in units of $e_0$. Employing these analytic
expressions the parameters of the model are determined by means of

\begin{description}

\item{i)} the electron-proton elastic differential cross sections measured by Sill et
al. \cite{sil:93} which overcome the preliminary results of \cite{arn:86} and are
available for $2.9 \leq Q^2 \leq 31.2 \hbox{GeV}^2$,

\item{ii)} the electric and magnetic form factors of the proton for $2.5 \leq
Q^2 \leq 8.83 \hbox{GeV}^2$ obtained by Bosted et al. \cite{bos:92} by means of a
Rosenbluth separation,

\item{iii)} the ratio $\sigma_n / \sigma_p$ of the $e n$ and $e p$ differential cross
sections extracted from the reaction $e d \rightarrow e p n$ at the quasi-elastic peak
\cite{roc:92} in the $Q^2$ range from $4 \hbox{GeV}^2$ to $10 \hbox{GeV}^2$,

\item{iv)} and both the neutron form factors at $Q^2 = 4 \hbox{GeV}^2$ extracted from
quasi-elastic $e-d$ cross sections via Rosenbluth separation \cite{lun:93}.

\end{description}

We want to emphasize that we only use data which are free from assumptions on
relationships between $G_E$ and $G_M$. A good fit to these data is achieved with the
following set of parameters:
\bes
f_S = 73.85 \hbox{MeV}, \; Q_S^2 = 3.22 \hbox{GeV}^2, \; a_S = 0.15, & & \nonumber \\ 
f_V = 127.7 \hbox{MeV}, \; Q_V^2 = 1.50 \hbox{GeV}^2, \; a_V = 0.05, & 
\kappa_V = 1.39, & c_1 = 5.8, \; c_2 = -12.5 . \nonumber \\ & & \label{param}
\ees

The results for proton and neutron electric and magnetic form factors are displayed in
Figs. 2 - 5. For comparison also shown are the predictions of two frequently cited
models, the vector meson dominance model of K\"orner and Kuroda \cite{koe:77}, and a
hybrid model \cite{gar:84} which combines the features of vector meson dominance
with QCD asymptotics. For general properties of the diquark model in connection with
nucleon form factors we refer to Ref. \cite{kro:91a}. Our results for the magnetic and
electric form factor of the proton match the data quite well 
\footnote{ The $G_M^p$ data of Ref. \cite{sil:93} have been extracted from elastic
$e-p$ cross sections under the assumption $G_M^p = (1+\mu_p) G_E^p$ which is not
satisfied in our model. The agreement with these data is only an indication that the
$e-p$ cross section is insignificantly affected by $G_E^p$.}.
The obvious difference between $G_M^p$ and $G_E^p$ signals that $F_2^p$ is by no means
small. This is an indication for sizable helicity flip transitions of the proton. 

For the neutron the situation seems not to be that satisfactory. However, we get the
correct sign and order of magnitude of the neutron magnetic form factor, anyhow a severe
challenge for any model based on perturbative QCD, and also our electric form factor
predictions exhibit the right trend at low $Q^2$. Missing quantitative agreement with
experiment should not be overvalued. Due to the small number of neutron data and their
relatively large error bars they are of minor importance for the $\chi^2$-value of our
fit. With the $Q^2$ values of the few data points available in the GeV region we
reach the limit of validity of the diquark model. On the other hand, the extraction of
these data from quasi-elastic $e-d$ cross sections is not completely straightforward and
cannot be performed model independent. Effects of final-state interactions and meson
exchange currents, not taken into account as yet, may still alter the results.
Therefore we did not attempt to achieve a better reproduction of the neutron data by
giving them an artificially high weight in the parameter-fitting procedure. 

In Fig.6 the four n-point contributions to the magnetic form factors $S_3$, $S_4$,
$V_3$, and $V_4$ are displayed to give some insight in their relative magnitudes.  
In order to see how they built up the magnetic form factors of the nucleon they have
to be combined according to Eqs.(\ref{gmp}) and (\ref{gmn}). 

At the end of this subsection a few words about the self-consistency of form factor
calculations within perturbative QCD are in order. As can be seen from Eqs.
(\ref{S3}) - (\ref{V4f}) the running QCD coupling constant $\alpha_S$ diverges in 
the end-point regions $x_1, y_1 \rightarrow 0,1$. 
The same happens in the pure quark HSP.
As a consequence, perturbation theory looses its self-consistency as a weak-coupling
expansion. One way out - actually the one we employ - is to \lq\lq freeze \rq\rq the
running coupling beyond an infrared cutoff \cite{cor:82}. In addition we make use of
distribution amplitudes which strongly suppress these dangerous kinematic regions.
The infrared cutoff represents a new scale within such an approach. The introduction
of this new scale can be avoided if the transverse momenta of the hadronic constituents
as well as radiative (Sudakov) corrections are taken into account. Radiative
corrections select components of the wave function with small spatial extend. The
numerical effect is similar to that of the infrared cutoff. This has been demonstrated
recently by Li and Sterman within the pure quark HSP for the form factors of the pion
\cite{li:92a} and and the nucleon \cite{li:92b}. They claim that in this way the
self-consistency of the perturbative treatment can be reestablished for 
$Q$ larger than $20\;\mbox{to}\;30\,\Lambda_{QCD}$ 
without introducing a scale in addition to $\Lambda_{QCD}$. We have
performed an analogous calculation for the S-diquark contribution to the proton form
factor and arrive at the same conclusion. A treatment of the endpoint regions in the
manner proposed by Li and Sterman diminishes the results presented above only by 
$10\,\mbox{to}\,20\%$. 
Since Sudakov corrections mainly depend on colour and not on spin we suppose
a similar behaviour for V diquarks. 
That small suppression can be compensated by adjusting 
 the parameters of the model appropriately. However this has not been
 done in a systematic way. We also note that there is further suppression
 due to the dependence of the baryon wave function on the intrinsic
 transverse momentum \cite{Jak:93}. We leave the systematic study
 of this effect and other higher twist contributions to a
 forthcoming paper.
\subsection{Hyperon form factors}
Employing the SU(6)-type flavour functions (\ref{flav3}) - (\ref{flav6}) we are
immediately in the position to extend the electromagnetic form-factor calculation to
other octet baryons. If one assumes the same DAs (\ref{phis}) and (\ref{phiv}) and
the same constants $f_S$ and $f_V$ for all members of the baryon octet no additional
ingredients are needed. A particular SU(6)-breaking scheme is already implied by the
different values of $f_S$ and $f_V$, the different functional form of $\Phi_S$ and
$\Phi_V$, and the mass dependence of the DAs (\ref{phis}), (\ref{phiv}). A similar
kind of SU(6) breaking proved also to be appropriate for the treatment of
proton-antiproton annihilation into hyperon antihyperon pairs within (a simplified
version of) the diquark model \cite{kro:89}. Although no data on electromagnetic
hyperon form factors are available at present, predictions for them might become of
interest in the future. The experiment proposed in \cite{gar:90} could provide data
with reasonable accuracy up to momentum transfers $Q^2 \approx 2 \hbox{GeV}^2$. Although this
$Q^2$ value is certainly not large enough for a direct comparison with our predictions,
the large-$Q^2$ trends provide perhaps some insight.

In terms of the four n-point functions already introduced in (\ref{S3}) - (\ref{V4}) the
magnetic form factors for $\Lambda$ and $\Sigma$ hyperons read:
\bes
G_M^{\Sigma^+} &=& e_u S_3 + e_{us} S_4 + (e_u  + 2 e_s) V_3 + (e_{us} + 2 e_{uu})
                 V_4  \, ,\\
G_M^{\Sigma^-} &=& G_M^{\Sigma^+} (u \leftrightarrow d) \, ,\\
2 G_M^{\Sigma^0} &=&  (e_u + e_d) S_3 +  (e_{us} + e_{ds}) S_4 \nonumber \\ & &
                  + (e_u + e_d + 4e_s) V_3 + (e_{us} + e_{ds} + 4 e_{ud}) V_4 \, ,\\
2 G_M^{\Lambda^0} &=& {1\over 3} (e_u + e_d + 4e_s) S_3 + 
                      {1\over 3} (e_{us} + e_{ds} + 4 e_{ud}) S_4 \nonumber \\ & &
                      + 3 (e_u + e_d) V_3 + 3 (e_{us} + e_{ds}) V_4 \, .
\ees
Similar expressions can be derived for $F_2$. From the charge factors in front of the
n-point functions it should be obvious whether the single quark is strange or
non-strange and correspondingly $150 \hbox{MeV}$ have to be added to the quark or diquark
mass occurring in the exponential of the DAs (\ref{phis}), (\ref{phiv}). Ignoring
this mass dependence one can look at Fig.6 to get a first estimate for the hyperon form
factors. One finds that
\be
G_M^{\Sigma^+} \approx G_M^p \; \; \hbox{and} \;\;  2 G_M^{\Sigma^0} \approx -
                                               2 G_M^{\Lambda^0}
                                              \approx - G_M^n \, .
\label{su6}
\ee
The magnetic form factors of $\Sigma^0$ and $\Lambda^0$ are small due to almost
perfect cancellation between the dominant contributions $S_3$ and $S_4$. The full
calculation, with the pertinent constituent masses in the DAs (\ref{phis}),
(\ref{phiv}) taken into account, confirms these observations as can be seen from
Fig.7. For comparison also shown is the result obtained by Chernyak et al.
\cite{cher:87} within the pure quark HSP (valid at high $Q^2$). Within the
theoretical uncertainties agreement between the diquark model and the HSP can be
observed.  At the lower $Q^2$-end there exist in addition predictions from a 
relativized constituent-quark model for $\Lambda^0$ and $\Sigma^0$ form factors 
\cite{war:91}. Also these results lie within the trend of the diquark model. 
At the end of this section we want to mention that in the SU(6)-limit the relations
(\ref{su6}) hold exactly. Therefore, the magnetic hyperon form factors could also give
us valuable hints at the amount of SU(6) breaking, in particular in the DAs of the
hyperons.
\section{THE AXIAL-VECTOR FORM FACTOR \hspace{20.0truecm} . OF THE NUCLEON}
Within the HSP the axial-vector form factor of the nucleon $g_A$ defined by
\be
\langle p; p_f, \lambda_f \vert \widehat{A^0}_\mu \vert n; p_i, \lambda_i \rangle =
i \cos \theta_c \bar{u} (p_f, \lambda_f) \left[ \gamma_\mu \gamma_5 g_A (Q^2) +
 (p_f - p_i)_\mu \gamma_5 f_A (Q^2) \right] u (p_i, \lambda_i)
\, , \label{wcurr}
\ee
is calculated along the same lines as the electromagnetic form factors. The same set
of Feynman diagrams has to be evaluated with the only difference that the photon is
replaced by a $W$-boson. No additional parameters enter the calculation
of $g_A$. Thus, it is an important touchstone for the whole picture and in
particular for the DA. Unfortunately data on $g_A$ are very scarce above $1 \hbox{GeV}^2$. The
only information comes from a measurement of the reaction $\nu_\mu n \rightarrow \mu^-
p$ in a deuterium target \cite{kit:83}. The analysis of the cross section is based
on the standard V-A theory. As a consequence of the PCAC hypothesis the contribution
from $f_A$ becomes negligible. Therefore the (unpolarized) differential cross section
suffices to determine $g_A$. Analogous to the electromagnetic form factors  and in
accordance with dimensional counting $g_A$ is usually parameterized as 
\be
g_A (Q^2) = {g_A (0) \over (1 + Q^2 / M_A^2 )^2} \, .
\label{gad}
\ee
From $\beta$-decay it is known that $g_A (0) = 1.23 \pm 0.01$. The experiment of
Kitagaki et al. \cite{kit:83}, with a maximum $Q^2$-value of $3 \hbox{GeV}^2$, gives $M_A
=1.05^{+0.12}_{-0.16} \hbox{GeV}$. The average over all $\nu$ data yields $M_A = 1.03 \pm
0.036 \hbox{GeV}$. But this result is dominated by low $Q^2$ data. Be that as it may, $Q^2$
is a bit small for application of perturbative QCD. In view of the results for
the magnetic form factor of the proton one may, however, hope that the dipole
parameterization (\ref{gad}) gives the right trend up to, say, $10$ or $15 \hbox{GeV}^2$.
If this is the case $Q^4 g_A$ is about $1.5^{+0.7}_{-0.9} \hbox{GeV}^4$ ($1.38 \pm 0.20$ from
the neutrino data) in that region.

This is not a severe constraint and causes no difficulty for the pure quark HSP. With
the DAs proposed in refs. \cite{che:84} and \cite{gar:87} Carlson et al.
\cite{car:87} found $Q^4 g_A = 1.36$ and $1.00 \hbox{GeV}^4$, respectively. Very
recently also Stefanis and Bergmann \cite{ste:93} found a similar result 
($Q^4 g_A = 1.44 \hbox{GeV}^4$ at $Q^2 \approx 10 \hbox{GeV}^2$ and for
$\Lambda_{QCD} = 180 \hbox{MeV}$) with a nucleon DA that combines features of the
Chernyak-Zhitnitsky DA \cite{che:84} and the Gari-Stefanis DA \cite{gar:87}.

Within the diquark model the calculation of $g_A$ is not as straightforward as in the
pure quark HSP. In the 4-point contributions the W interacts with diquarks and has to
change their flavour and in some cases even their spin (e.g.: $W^+ S_{[u,d]}
\rightarrow V_{\{u, u\} }$). Taking into account such transitions would require the
introduction of new diquark form factors. One may, however, argue that the 4-point
contributions are negligible since diquarks break up when they interact with a W. This
would mainly lead to particle production. We assume, therefore, the dominance of
the 3-point contribution and obtain the simple result
\be
g_A = S_3 -  V_3 \, .
\ee
From Fig.6 it is clear that our prediction for $g_A$ agrees reasonably well with the
parameterization (\ref{gad}). For $Q^2 = 15 \hbox{GeV}^2$ we get $Q^4 g_A = 1.15 \hbox{GeV}^4$.
\section{SUMMARY}
In this paper we have presented a new parameterization of the quark-diquark model of
baryons. This model was introduced originally in Refs. \cite{kro:91a} - \cite{kro:93} in
order to describe exclusive hadronic reactions at intermediate momentum transfers. In
\cite{kro:91a} the parameters of the model were determined by means of elastic
electron-nucleon scattering data. A refit of the parameters became necessary due to the
publication of new and more accurate electron-nucleon data \cite{bos:92}, \cite{sil:93},
\cite{roc:92}, \cite{lun:93}. With the new set of parameters good agreement with the
electric and magnetic form-factor data of the proton for $Q^2 \ageq 4 \hbox{GeV}^2$ has
been achieved. We have not attempted to optimize our results concerning the neutron
form factors, since corresponding data are still very sparse and only known for $Q^2
\leq 4\hbox{GeV}^2$ - the lower limit of validity of the diquark model. Nevertheless,
the diquark-model results for the electromagnetic neutron form factors lie within the
high $Q^2$ trend indicated by the existing experiments. More and better data for the
neutron form factors, say between $4$ and $20\hbox{GeV}^2$, would certainly be very
helpful and could furthermore facilitate the decision between the different theoretical
models applicable in that kinematic region. 

In the spirit of Li and Sterman \cite{li:92a}, \cite{li:92b} we have also investigated
the influence of transverse momentum and Sudakov corrections for the S-diquark part of
our model. It turned out that the suppression of the dangerous end-point regions $x
\rightarrow 0, 1$ by means of such corrections is already implemented in our model in 
an effective way through the restriction $\alpha_s \leq 0.5$ on the strong
coupling constant and the particular choice of the DAs. Since Sudakov corrections
mainly depend on colour and not on spin we suppose them also to be of minor importance
for V-diquarks.

In addition to electromagnetic nucleon form factors, we have considered the
axial-vector form factor of the nucleon and electromagnetic form factors of $\Lambda$
and $\Sigma$ hyperons. Under the assumption that contributions in which the W boson
couples to the diquarks are negligible -- due the necessary change of flavour the
diquarks tend to break up -- the diquark-model predictions for $g_A$ compare reasonably
well with the dipole parameterization of $g_A$ extracted from $\nu_\mu n \rightarrow
\mu^- p$ data \cite{kit:83}. For the magnetic $\Lambda$ and $\Sigma^0$ form factors
the diquark model interpolates between the corresponding results of the pure quark
HSP \cite{cher:87}, valid at high $Q^2$ and predictions of a constituent quark model
\cite{war:91}, valid at small $Q^2$. In the limit of exact $SU(6)$ spin-flavour symmetry
the hyperon form factors are related by $G_M^{\Sigma^+} = G_M^p$ and 
$2 G_M^{\Sigma^0}= - 2 G_M^{\Lambda^0} = - G_M^n$. 
Since we have used flavour dependent DAs our results
deviate from these relations. Data for these hyperon form factors could give us
valuable clues at the amount of $SU(6)$ breaking and the flavour dependence of
octet-baryon DAs.

Finally it should be mentioned that there exist already other successful applications of
the new diquark-model parameterization presented here. These include the time-like form
factor of the proton, $\gamma \gamma \rightarrow p \bar{p}$, $\eta_c \rightarrow p
\bar{p}$ \cite{kro:93b}, and (virtual) Compton scattering off protons \cite{schue:93}.
\ns
\ns
\section{FIGURE CAPTIONS}
\begin{description}

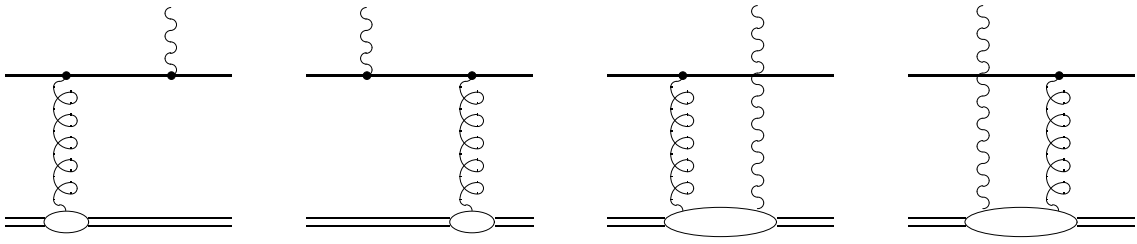
\item[Fig.1] Diagrams contributing to electromagnetic and weak form factors of baryons.
The double lines represent either an S or a V diquark.

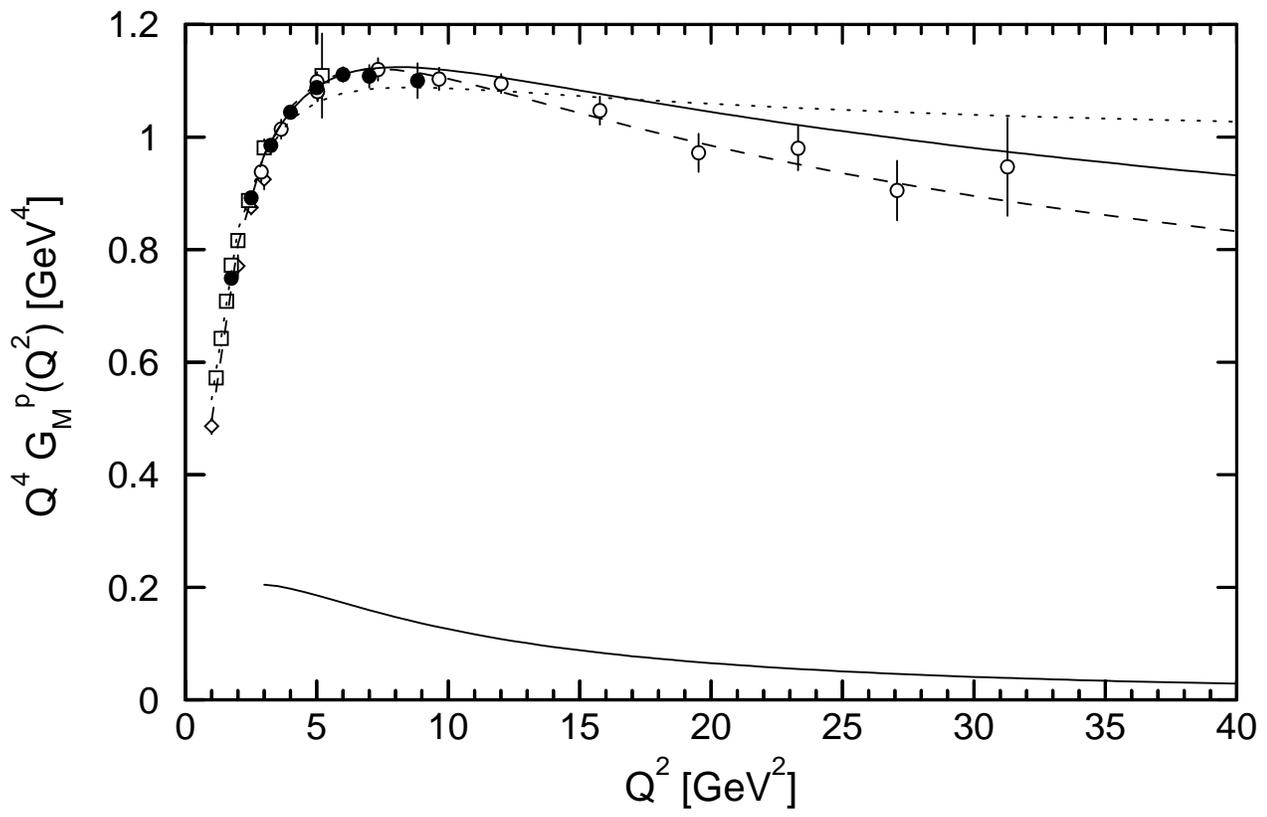
\item[Fig.2] The magnetic form factor of the proton. Shown are the results of
the diquark model (solid line), the vector-meson-dominance model \cite{koe:77}
(dotted line), and a hybrid model \cite{gar:84} (dashed line). Data are taken from
\cite{sil:93} ($\circ$), \cite{bos:92} ($\bullet$), \cite{hoe:76} ($\square{7pt}$), and
\cite{wal:89} ($\diamond$). The lower solid line shows the Pauli form factor $F_2^p$
as resulting from the diquark model.

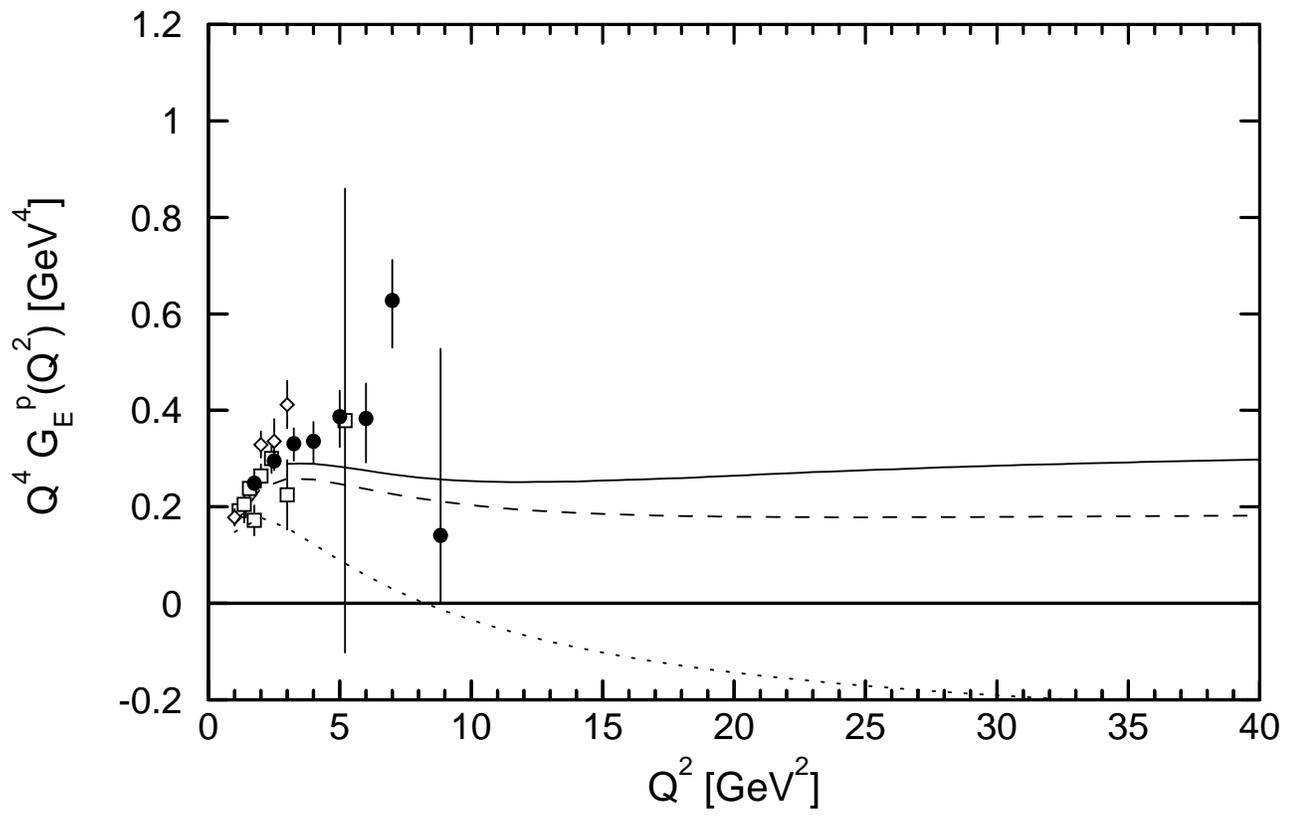
\item[Fig.3] The electric form factor of the proton. Data are taken from \cite{bos:92} 
($\bullet$), \cite{hoe:76} ($\square{7pt}$), and \cite{wal:89} ($\diamond$). Curves
same as in Fig.2.

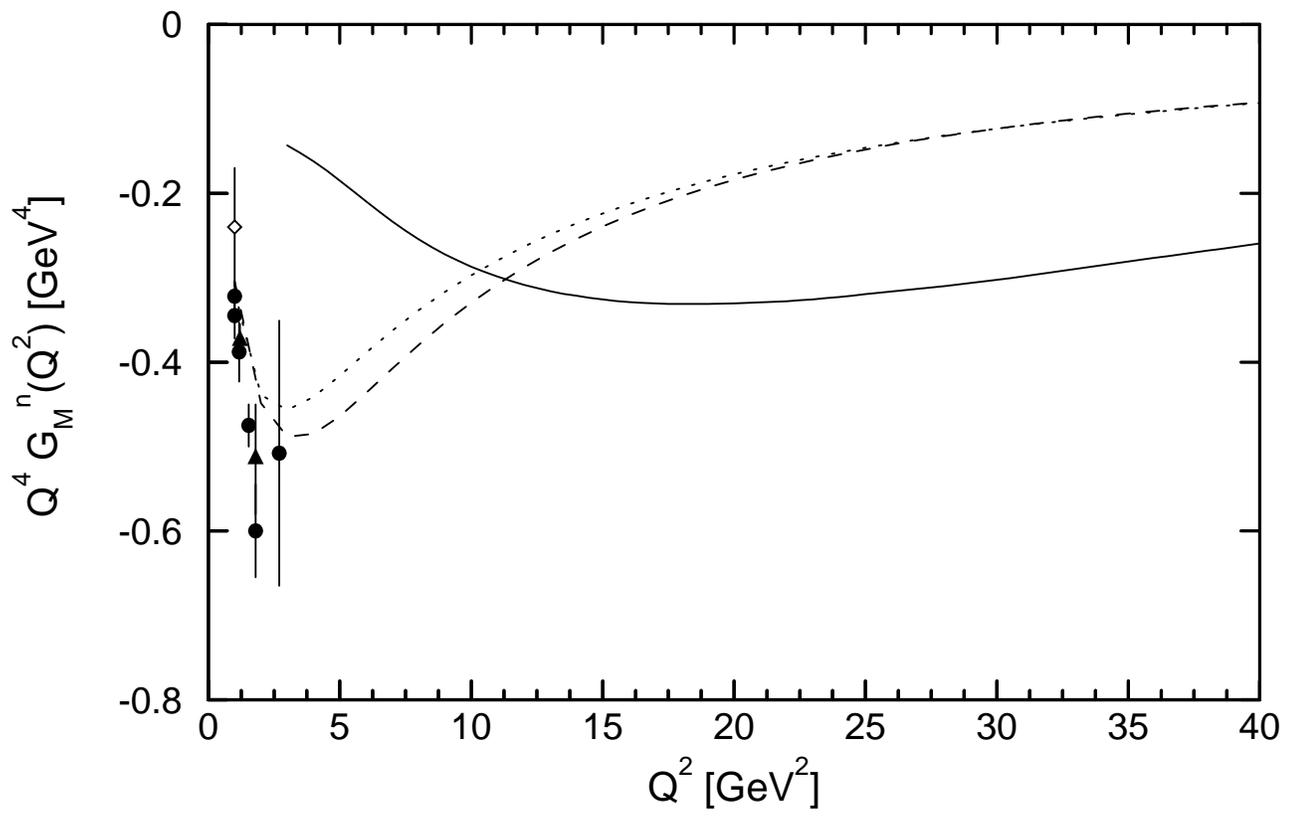
\item[Fig.4] The magnetic form factor of the neutron. Data are taken from
\cite{ake:64} ($\triangle$), \cite{han:73} ($\diamond$), \cite{bar:73} ($\bullet$),
and \cite{lun:93} ($\circ$). Curves same as in Fig.2. 

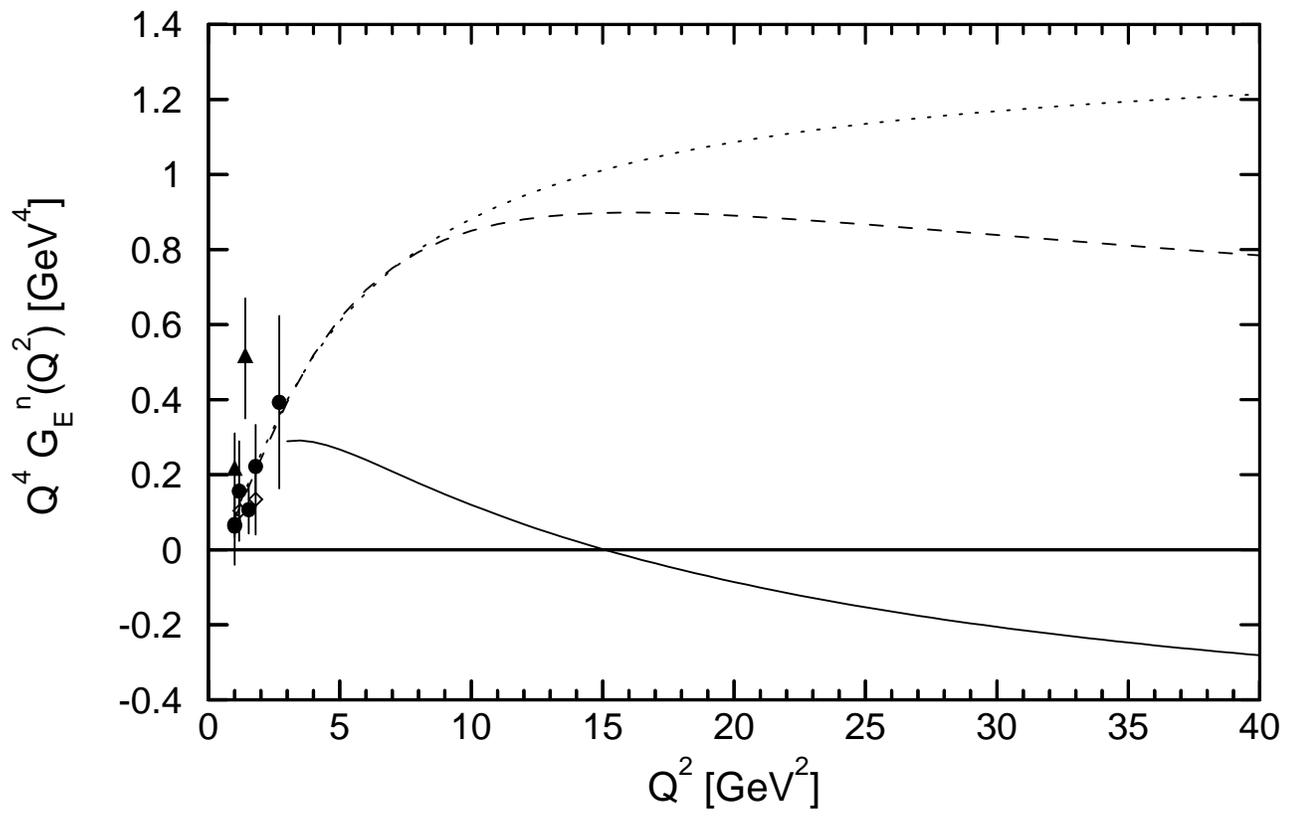
\item[Fig.5] The electric form factor of the neutron. Data are taken from
\cite{ake:64} ($\triangle$), \cite{han:73} ($\diamond$), \cite{bar:73} ($\bullet$),
and \cite{lun:93} ($\circ$). Curves same as in Fig.2.  

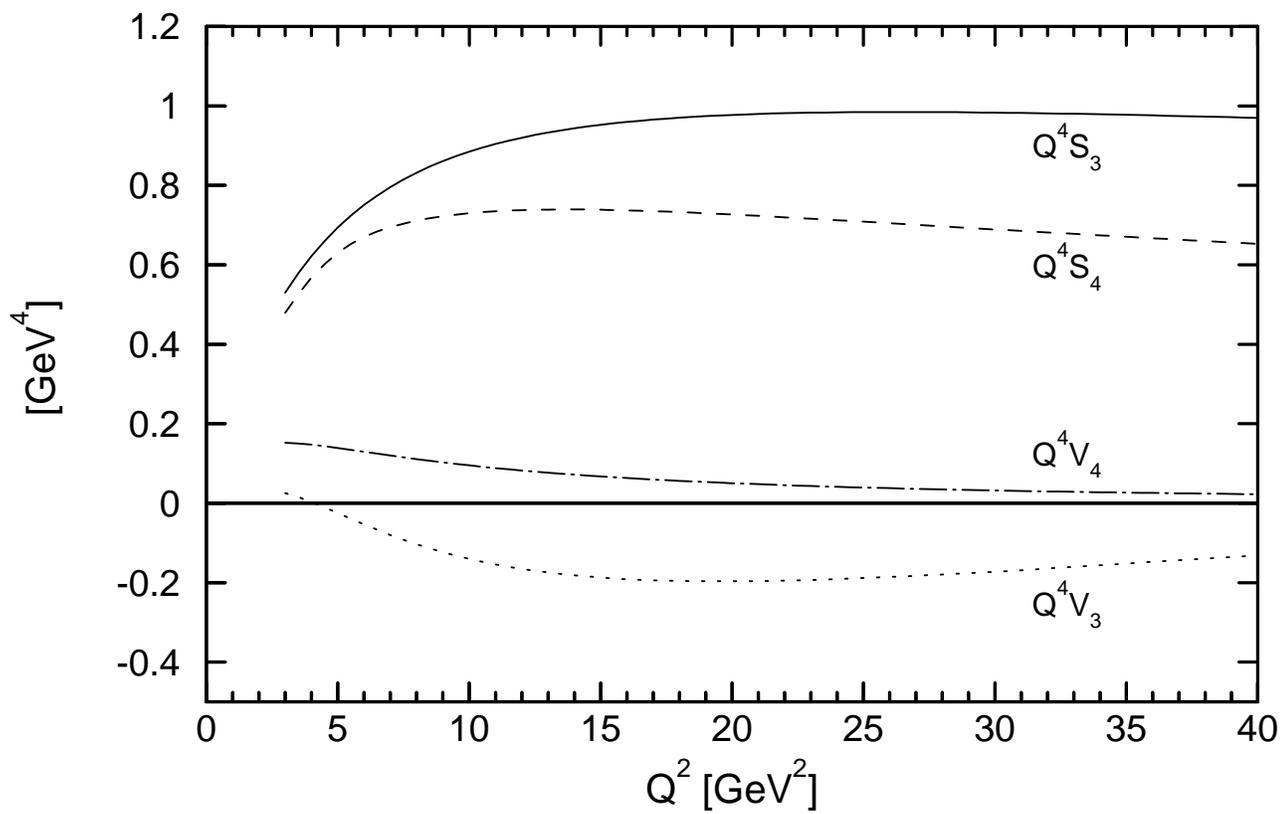
\item[Fig.6] The 3- and 4-point terms contributing to the magnetic form factors of the
nucleon (cf. (\ref{gmp}) and (\ref{gmn})).

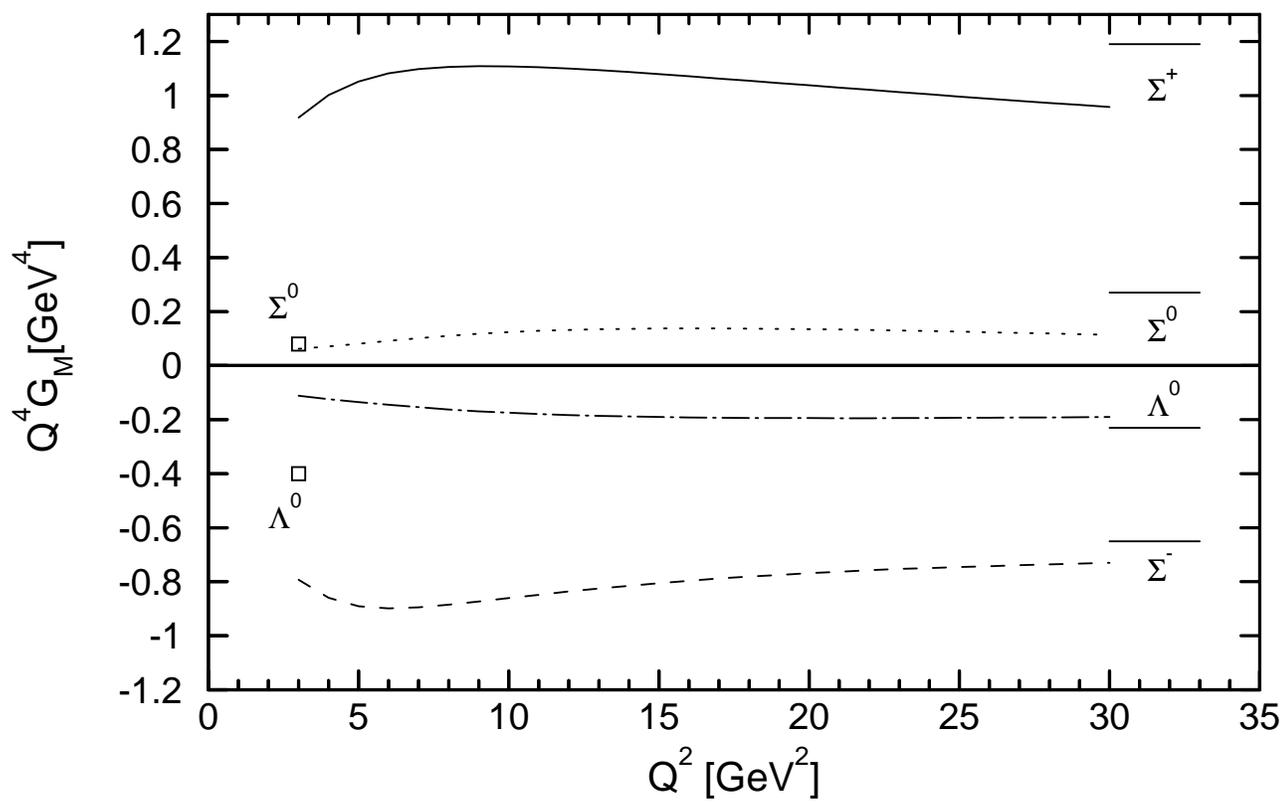
\item[Fig.7] Our predictions for the magnetic form factors of $\Lambda$ and $\Sigma$
             hyperons. The dashes and squares represent the corresponding results from
             the pure quark HSP \cite{cher:87} and a relativized constituent quark model
             \cite{war:91}, respectively. 
 
\end{description}
\end{document}